\documentclass{article}

\usepackage{ml4cps}
\renewcommand{\crowncopyrightyear}{2026}

\usepackage[utf8]{inputenc} 
\usepackage[T1]{fontenc}    
\usepackage{hyperref}       
\usepackage{url}            
\usepackage{booktabs}       
\usepackage{amsfonts}       
\usepackage{nicefrac}       
\usepackage{microtype}      
\usepackage{lipsum}         
\usepackage{graphicx}
\usepackage{doi}
\usepackage{amsmath}
\usepackage{cleveref}       
\usepackage{wrapfig}
\usepackage{siunitx}
\usepackage{subcaption}
\usepackage{pdflscape}
\usepackage{multirow}
\usepackage{placeins}
\usepackage{threeparttable}
\usepackage[
  backend=biber,
  style=numeric-comp,
  sorting=none,
  giveninits=true,
  maxbibnames=99,
  doi=true,
  url=true,
  eprint=true
]{biblatex}

\title{Scale-aware and spatially-localised optimisation of kilometre-scale MLWP models}

\date{}

\newif\ifuniqueAffiliation

\ifuniqueAffiliation 
\author{ \href{https://orcid.org/0000-0003-2997-0150}{\includegraphics[scale=0.06]{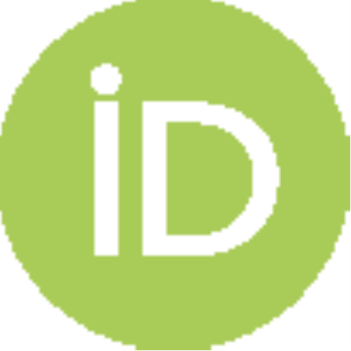}\hspace{1mm}James Chappell}\\
	Met Office\\
	UK\\
	\texttt{james.chappell@metoffice.gov.uk} \\
	\And
	\href{https://orcid.org/0000-0000-0000-0000}{\includegraphics[scale=0.06]{figures/orcid.png}\hspace{1mm}Second Author} \\
	Affiliation\\
	Address\\
	\texttt{author@mail.com} \\
}
\else
\usepackage{authblk}

\newbox{\orcid}\sbox{\orcid}{\includegraphics[scale=0.06]{figures/orcid.png}} 
\author[1]{%
    \href{https://orcid.org/0000-0003-2997-0150}{%
        \usebox{\orcid}\hspace{1mm}James Chappell%
    }%
    \thanks{Corresponding author:
        \href{mailto:james.chappell@metoffice.gov.uk} {\texttt{james.chappell@metoffice.gov.uk}}%
    }%
}
\author[2]{%
	\href{https://orcid.org/0000-0001-9708-7058}
        {\usebox{\orcid}\hspace{1mm}Sophie Luise Arana%
    }%
    \thanks{Current address: LGND AI, Inc., California, USA}%
}
\author[1]{%
    \href{https://orcid.org/0000-0002-4325-6685}
    {\usebox{\orcid}\hspace{1mm}Jack Bowyer}%
}
\author[1]{%
	{Matthew Burnand}%
}
\author[2]{%
	\href{https://orcid.org/0000-0002-9264-1539}{\usebox{\orcid}\hspace{1mm}Alejandro Coca-Castro}%
}
\author[2]{%
	\href{https://orcid.org/0000-0003-0321-009X}{\usebox{\orcid}\hspace{1mm}Evangeline Corcoran}%
}
\author[2]{%
	\href{https://orcid.org/0000-0002-8499-0720}{\usebox{\orcid}\hspace{1mm}Eric G. Daub}%
}
\author[1]{%
	\href{https://orcid.org/0009-0007-4756-8751}{\usebox{\orcid}\hspace{1mm}Tom Dunstan}%
}
\author[2]{%
	\href{https://orcid.org/0009-0006-0028-2871}{\usebox{\orcid}\hspace{1mm}Noushin Eftekhari}%
}
\author[1]{%
	\href{https://orcid.org/0000-0002-6805-1379}{\usebox{\orcid}\hspace{1mm}Anna-Louise Ellis}%
}
\author[2]{%
	\href{https://orcid.org/0000-0002-3646-3504}{\usebox{\orcid}\hspace{1mm}J. Scott Hosking}%
}
\author[2]{%
	\href{https://orcid.org/0009-0003-1974-3162}{\usebox{\orcid}\hspace{1mm}Manvendra Janmaijaya}%
}
\author[1]{%
	{Sam Madge}%
}
\author[2]{%
	\href{https://orcid.org/0000-0001-8892-2410}{\usebox{\orcid}\hspace{1mm}David Salvador-Jasin}%
}
\author[1]{%
	\href{https://orcid.org/0000-0002-7419-0789}{\usebox{\orcid}\hspace{1mm}Ben Shipway}%
}
\author[2]{%
        {Nathan Simpson%
    }%
    \thanks{Current address: AI for Physics, Mistral, London, United Kingdom}%
}
\author[2]{%
    \href{https://orcid.org/0000-0002-8177-5582}{\usebox{\orcid}\hspace{1mm}Oliver Strickson}%
}
\author[1]{%
	\href{https://orcid.org/0000-0003-0383-7378}{\usebox{\orcid}\hspace{1mm}David Walters}%
}
\affil[1]{Met Office, Exeter, United Kingdom}
\affil[2]{The Alan Turing Institute, London, United Kingdom}
\fi

\renewcommand{\shorttitle}{Scale-aware and spatially-localised optimisation of km-scale MLWP models}

\hypersetup{
pdftitle={contribution title},
pdfsubject={contribution subject},
pdfauthor={First author, Second author},
pdfkeywords={First keyword, Second keyword, Third keyword},
}

\begin{document}

\maketitle

\begin{abstract} 
Preserving fine-scale structure without degrading spatial accuracy is a central challenge for kilometre-scale machine-learning weather prediction (MLWP). We introduce a loss based on the dual-tree complex wavelet transform that constrains atmospheric fields jointly by scale, location, and orientation. We evaluate its performance using an experimental joint global--regional FastNet configuration, which produces \SI{1.5}{km} forecasts over the UK alongside $\sim\SI{31}{km}$ global forecasts. Relative to both mean-squared error (MSE) and an FFT-based spectral objective, the wavelet loss substantially improves spectral fidelity. Compared with MSE training, it maintains similar pointwise performance, improves neighbourhood-based verification, and produces sharper precipitation fields with more realistic intensity distributions. These results demonstrate that spatially localised wavelet constraints can preserve multiscale atmospheric structure while largely avoiding the loss of pointwise accuracy associated with globally supported spectral objectives. 
\end{abstract}

\section{Introduction}

Machine-learning weather prediction (MLWP) models can produce accurate forecasts at a fraction of the computational cost of conventional numerical weather prediction~\cite{keisler2022,pathak2022,Bi2023,lam2023,Chen2023,lang2024,daub2025,dunstan2026}. As these models are extended to higher spatial resolutions~\cite{Bodnar2025,goecke2026}, a variety of regional MLWP approaches have emerged using stretched-grid~\cite{2026nipen,nordhagen2025,Wijnands2026}, limited-area~\cite{oskarsson2023,adamov2025,Xu2025,larsson2025}, and dataset-specific encoder--decoder architectures (as used in this study). At kilometre-scale resolutions, the training data capture local weather systems and spatially varying surface influences that are poorly represented by coarser atmospheric fields, providing information important for forecasting high-impact weather and issuing local warnings~\cite{clark2016}. However, resolving these features in the training data does not guarantee their preservation in forecasts, regardless of model architecture.

\begin{figure}[t]
    \centering
    \includegraphics[width=1\linewidth]{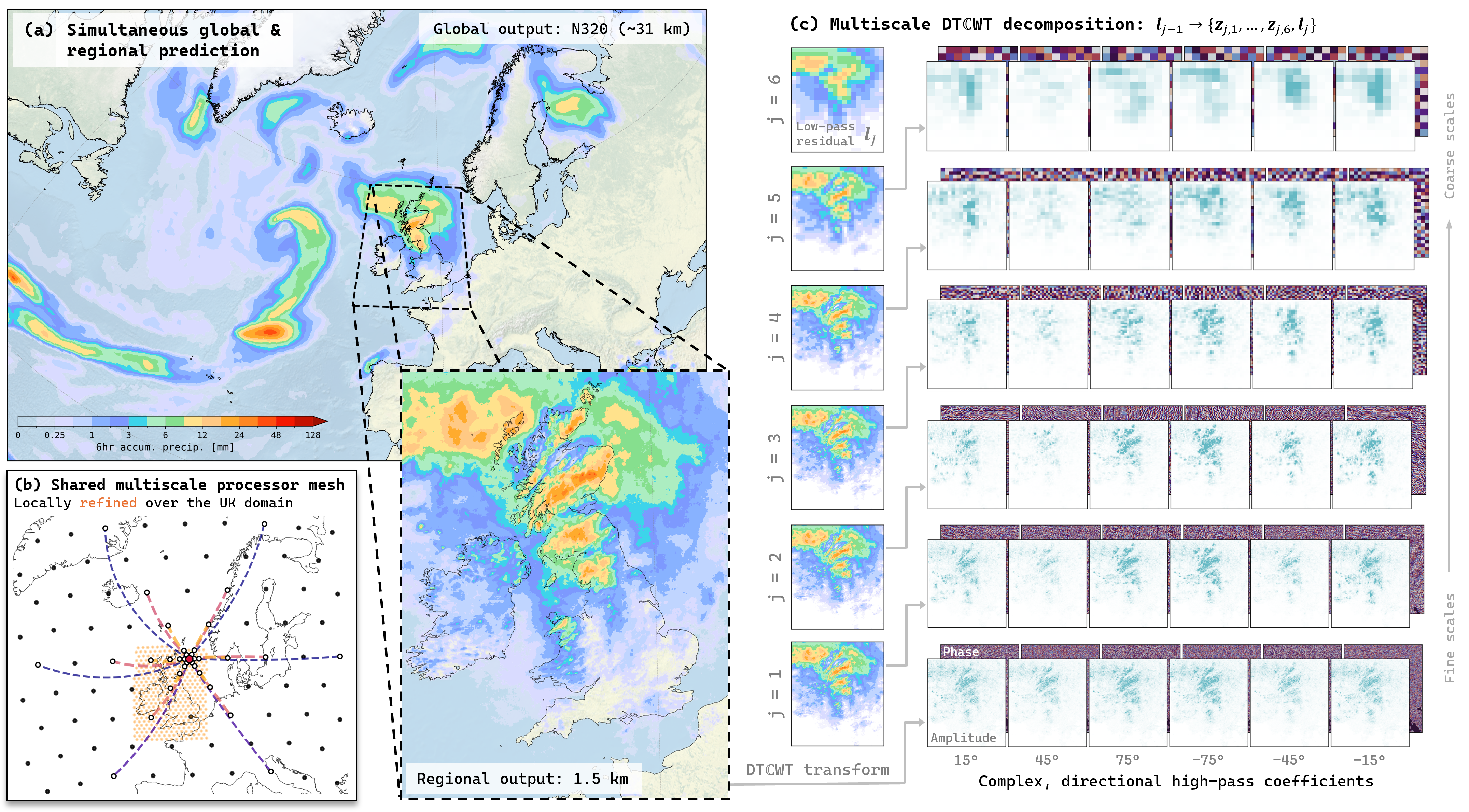}
    \caption{ 
    Overview of the regional MLWP architecture and wavelet training objective. (a) The model produces simultaneous forecasts on a global N320 reduced Gaussian grid and a \SI{1.5}{km} regional grid centred on the UK; the example shows \SI{6}{hour} accumulated precipitation valid at 2023-09-28T00Z UTC at a lead time of \SI{24}{hours}. (b) The shared latent processor mesh spans the global domain and is locally refined over the regional domain. Black and orange points denote global and regional processor nodes, respectively, while curves show representative multiscale processor edges from one regional node. The edges are coloured according to their length and are represented by arcs for visual clarity. (c) During loss calculation, the prediction and target are decomposed using a $J$-level DT$\mathbb{C}$WT. At each level, the preceding low-pass component is separated into a coarser low-pass component and six complex directional wavelet-bands. The high-pass coefficients contribute amplitude- and coherence-sensitive losses that can be weighted independently by scale, while the final low-pass residual is compared using MSE. }
    \label{fig:overview}
\end{figure}

This challenge is particularly acute for deterministic models trained using pointwise mean-squared error (MSE). Kilometre-scale atmospheric fields contain intermittent and sharply localised structures, such as convective precipitation, whose precise evolution and location are difficult to predict. Because MSE strongly penalises spatial displacement errors, minimising this objective tends to favour smooth predictions over uncertain fine-scale structures, suppressing small-scale variability and reducing effective resolution~\cite{subich2025}. While probabilistic approaches can represent uncertain fine-scale variability through ensembles of plausible realisations~\cite{alet2025,bonev2025,Lang2026b}, deterministic models provide a computationally efficient and controlled setting for investigating how training objectives influence forecast structure.

One approach to mitigate smoothing of deterministic forecasts is incorporating spectral information into the training objective. Subich et al.~\cite{subich2025} introduced the adjusted mean-squared error (AMSE), which decomposes forecast errors spectrally and separates contributions associated with amplitude and coherence. Originally formulated using spherical harmonics for global models, AMSE can be adapted to regional domains using Fourier transforms, while related amplitude-based spectral losses have shown promise in probabilistic regional forecasting~\cite{nordhagen2025}. However, their globally supported basis functions limit their ability to constrain forecast errors locally in space. Improvements in spectral fidelity may therefore come at the expense of accurately positioning individual features, potentially degrading pointwise verification metrics such as root-mean-squared error (RMSE)~\cite{dunstan2026}. This motivates objectives that preserve fine-scale variability while retaining sensitivity to the location of forecast structures.

Here, we develop a deterministic loss based on the two-dimensional dual-tree complex wavelet transform (DT$\mathbb{C}$WT)~\cite{kingsbury1998,selesnick2005}, illustrated schematically in Figure~\ref{fig:overview}. By decomposing fields jointly by scale, location, and orientation, the DT$\mathbb{C}$WT allows loss contributions to be calculated and weighted independently across spatial scales. The approximate shift invariance~\cite{kingsbury1999} and directional selectivity of the DT$\mathbb{C}$WT make it well suited to localised atmospheric structures, while its complex coefficients support amplitude- and coherence-sensitive terms analogous to those used in spectral AMSE. Wavelets have previously been applied to atmospheric analysis and verification~\cite{scovell2020,buschow2021}, with recent machine-learning applications focusing primarily on precipitation, including generative downscaling~\cite{yi2026} and a composite DT$\mathbb{C}$WT--Fourier loss for global prediction~\cite{An2026}. Our standalone objective extends wavelet-based optimisation to multivariable, kilometre-scale forecasting, aiming to preserve fine-scale spectral variability without compromising spatial accuracy. The underlying multiscale decomposition could also provide a foundation for scale-aware probabilistic training objectives, complementing recent developments in probabilistic MLWP~\cite{lang2025,lang2026}.

We evaluate the proposed objective using a joint global--regional configuration developed as part of ongoing work towards a regional version of the medium-range MLWP model FastNet~\cite{daub2025,dunstan2026}. The model simultaneously produces kilometre-scale forecasts over the UK and coarser forecasts globally, providing a controlled setting in which to compare wavelet-based training with MSE and an FFT-based spectral objective. We show that the wavelet loss preserves fine-scale spectral energy while maintaining strong spatial localisation of the training gradient. It achieves pointwise performance comparable to MSE, outperforms MSE under neighbourhood-based verification, and improves spectral fidelity relative to both MSE and the purely spectral objective.

\section{Deterministic wavelet loss}
\label{sec:wavelet_loss}

To combine scale-selective amplitude matching with greater spatial localisation, we formulate the loss in the dual-tree complex wavelet domain. Let $\widehat{\boldsymbol{x}}$ and $\boldsymbol{x}$ denote predicted and target fields, respectively, on a two-dimensional input grid. Applying a $J$-level DT$\mathbb{C}$WT to a field $\boldsymbol{x}$ produces complex directional band-pass coefficient fields, $\boldsymbol{z}_{j,o}[\boldsymbol{x}]$, at each level $j=1,\ldots,J$ and at six approximate orientations ($\pm\SI{15}{\degree}$, $\pm\SI{45}{\degree}$, and $\pm\SI{75}{\degree}$), together with a final real-valued low-pass residual, $\boldsymbol{l}_J[\boldsymbol{x}]$: 

\begin{equation} 
\mathcal{W}\!\left[\boldsymbol{x}\right] = \left( \boldsymbol{l}_{J}[\boldsymbol{x}], \left\{ \boldsymbol{z}_{j,o}[\boldsymbol{x}] \right\}_{j=1,\ldots,J;\;o=1,\ldots,6} \right). 
\label{eq:dtcwt_outputs} 
\end{equation} 

We write $\widehat{\boldsymbol{z}}_{j,o} =\boldsymbol{z}_{j,o}[\widehat{\boldsymbol{x}}]$ and $\boldsymbol{z}_{j,o} =\boldsymbol{z}_{j,o}[\boldsymbol{x}]$ for the prediction and target coefficients, respectively, with analogous notation for the low-pass residuals.

The index $j=1$ denotes the finest level, with increasing $j$ corresponding to progressively coarser scales. Because the transform downsamples the coefficient fields at successive levels, each level has its own coefficient grid and set of spatial indices, $\mathcal{S}_j$. Thus, $\boldsymbol{r}\in\mathcal{S}_j$ denotes a location on the level-$j$ grid rather than on the original input grid.

For each high-pass scale and orientation, we calculate the band-integrated wavelet powers of the prediction and target: 

\begin{equation} 
\widehat{P}_{j,o} = \sum_{\boldsymbol{r}\in\mathcal{S}_{j}} \left| \widehat{z}_{j,o}(\boldsymbol{r}) \right|^2, \qquad P_{j,o} = \sum_{\boldsymbol{r}\in\mathcal{S}_{j}} \left| z_{j,o}(\boldsymbol{r}) \right|^2. 
\label{eq:wavelet_band_power} 
\end{equation} 

Following the AMSE formulation~\cite{subich2025}, the loss for each wavelet band combines an amplitude error with a coherence-dependent penalty: 

\begin{equation} 
\mathcal{L}_{j,o} = \left( \sqrt{\widehat{P}_{j,o}} - \sqrt{P_{j,o}} \right)^2 + 2\max\left(\widehat{P}_{j,o},P_{j,o}\right) \left(1-\rho_{j,o}\right), 
\label{eq:wavelet_band_loss} 
\end{equation} 

where the normalised coherence is 

\begin{equation} 
\rho_{j,o} = \frac{ \displaystyle \sum_{\boldsymbol{r}\in\mathcal{S}_{j}} \operatorname{Re}\left[ \widehat{z}_{j,o}(\boldsymbol{r}) z_{j,o}^{*}(\boldsymbol{r}) \right] }{ \left(\widehat{P}_{j,o}P_{j,o}\right)^{1/2} }. 
\label{eq:wavelet_band_coherence} 
\end{equation} 

The wavelet-band loss is summed over the six orientations and combined across scales using configurable weights, $\alpha_j$. The final low-pass residual represents the coarsest spatial structure and is compared directly using MSE. The total loss is 

\begin{equation} \mathcal{L}_{\mathrm{wav}} = \frac{1}{N} \left[ \sum_{j=1}^{J}\alpha_j \sum_{o=1}^{6}\mathcal{L}_{j,o} + \left\| \widehat{\boldsymbol{l}}_{J} - \boldsymbol{l}_{J} \right\|_2^2 \right], 
\label{eq:wavelet_total_loss} 
\end{equation} 

where $N$ is the number of points in the original input grid. 

The two terms in Equation~\ref{eq:wavelet_band_loss} measure distinct aspects of the error within each scale--orientation band. The amplitude term compares the band-integrated powers of the prediction and target, while the second term applies a power-weighted penalty for disagreement between their complex coefficient patterns. Prediction and target coefficients are paired at corresponding locations in the coherence numerator before their cross-power is summed, making the loss sensitive to their relative spatial arrangement as well as their band-integrated powers. The loss therefore distinguishes forecasts with similar band-integrated power but different spatial organisation.

Although Equation~\ref{eq:wavelet_total_loss} is additive across levels and orientations, each band loss is computed from statistics aggregated over all locations in $\mathcal{S}_j$ and is therefore not a sum of independent coefficient-wise losses---the corresponding locations and gradients remain coupled within each band. Moreover, each coefficient is the response of a compactly supported wavelet filter to a neighbourhood of input values, rather than a pointwise value on the original grid. Back-propagating through these filters consequently produces more spatially localised field-space gradients than transforms with globally supported basis functions, such as the Fourier transform, as shown in Appendix~\ref{app:spatial_attribution}. The following subsection demonstrates how the separate level-wise terms provide scale-selective loss contributions and field-space gradients.

\subsection{Scale selectivity} 
\label{sec:scale_selectivity} 

\begin{figure}[ht]
    \centering
    \includegraphics[width=1\linewidth]{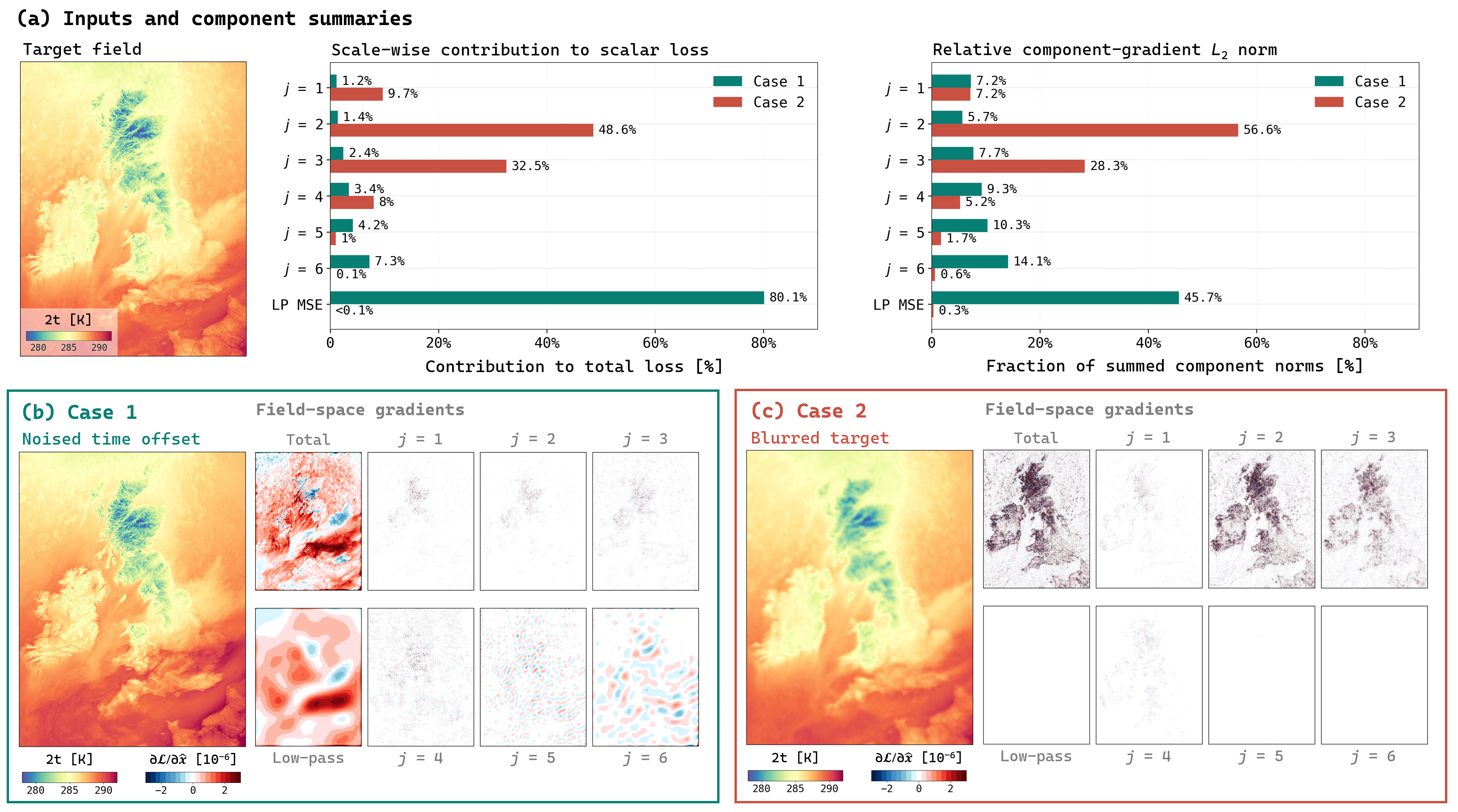}
    \caption{ Scale-wise decomposition of the wavelet loss for two constructed prediction fields. (a) Target field and component-wise summaries for a perturbed \SI{6}{hour} time-offset prediction (Case 1) and a Gaussian-blurred target with $\sigma=3$ grid cells (Case 2). The scalar-loss percentages give the exact contribution of the final low-pass term and each high-pass wavelet level to the total loss. The gradient percentages give each component-gradient $L_2$ norm as a fraction of the sum of the component-gradient $L_2$ norms; they therefore provide a relative measure of gradient magnitude rather than an additive decomposition of the norm of the total gradient. (b,c) The two constructed predictions and the gradients of the total loss and its low-pass and level-wise components with respect to the predicted field on the original input grid. Each map shows the signed first-order sensitivity of the corresponding scalar loss component and therefore the field-space training signal supplied by that component.} 
    \label{fig:wavelet_decomposition}
\end{figure}

Figure~\ref{fig:wavelet_decomposition} illustrates the scale selectivity of the wavelet loss using two contrasting constructed prediction fields designed to probe different error regimes. Case 1 is the atmospheric state \SI{6}{hours} before the target time with Gaussian noise added ($\sigma=5\%$), producing errors across a broad range of spatial scales. Case 2 is obtained by applying a Gaussian blur with $\sigma=3$ grid cells to the target, preserving its large-scale structure while selectively attenuating finer-scale variability. Uniform scale weights, $\alpha_j=1$, are used in both cases.

The scalar loss is decomposed exactly into the wavelet-band components, $\mathcal{L}_{j} =N^{-1}\alpha_j\sum_{o=1}^{6}\mathcal{L}_{j,o}$, and the final low-pass component, $\mathcal{L}_{\mathrm{LP}} =N^{-1}\|\widehat{\boldsymbol{l}}_J-\boldsymbol{l}_J\|_2^2$. For each component, we also calculate its gradient with respect to the predicted field on the original input grid, $\nabla_{\widehat{\boldsymbol{x}}}\mathcal{L}_{j}$, by back-propagating through the wavelet transform. These field-space gradients represent the signed first-order sensitivity of each scalar component to changes in the predicted field, revealing the scale and spatial structure of the training signal presented to the forecasting model. The total field-space gradient is the sum of the low-pass and wavelet-band gradients, although the eventual parameter update also depends on back-propagation through the model and on the optimiser.

The two cases produce clearly different responses. For the perturbed time-offset prediction (Case 1, Fig.~\ref{fig:wavelet_decomposition}b), broad structural errors dominate the scalar objective, with the low-pass component contributing $80\%$ of the total loss. Its share of the summed component-gradient norms is much lower, at $46\%$, with the remaining gradient magnitude distributed across the wavelet-band levels and increasing towards the coarser levels. The wavelet components therefore provide substantial scale-dependent training signals even when the scalar loss is dominated by its low-pass component. 

By contrast, the blurred prediction (Case 2, Fig.~\ref{fig:wavelet_decomposition}c) retains the target's large-scale structure, giving a negligible low-pass contribution to both the loss and gradient magnitude. Its response is concentrated at wavelet-bands $j=2$ and $j=3$, which together account for approximately $81\%$ of the scalar loss and $85\%$ of the summed component-gradient norms. This concentration reflects both the range of scales attenuated by the Gaussian filter and the distribution of variance in the target field: although the finest scales undergo strong fractional attenuation, the \SI{2}{m} temperature field contains more variance within the affected wavelet-bands at $j=2$ and $j=3$ than at $j=1$.

Together, these examples show that the wavelet objective produces scale-dependent scalar losses and field-space training signals. Broad structural errors activate the low-pass and coarser-scale terms, whereas selectively removing finer-scale variability concentrates the response in the affected fine-scale wavelet-band components. The spatial localisation of the complete wavelet-loss gradient is compared with that produced by MSE and FFT-AMSE in Appendix~\ref{app:spatial_attribution}.

\subsection{Evolution of scale-wise loss contributions during training} 
\label{sec:scale_loss_evolution} 

\begin{figure}[ht]
    \centering
    \includegraphics[width=1\linewidth]{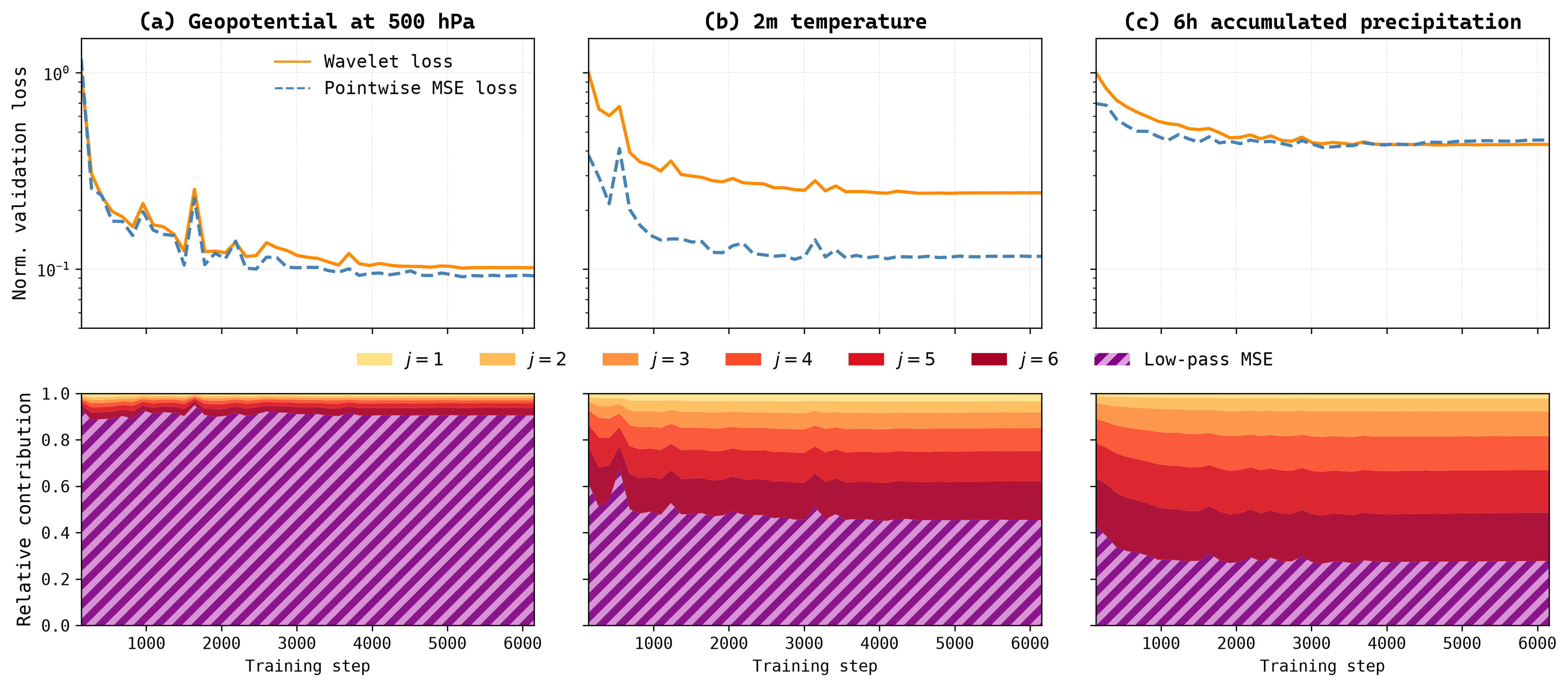}
    \caption{ 
    Evolution of validation losses and relative scale-wise wavelet-loss contributions over the first 25\% of Stage~3a training for (a) \SI{500}{hPa} geopotential, (b) \SI{2}{m} temperature, and (c) 6h accumulated total precipitation. All quantities are evaluated over the regional UKV domain using validation forecasts from the wavelet-trained model; the global ERA5 component is not included. Upper panels show pointwise MSE and the complete wavelet objective, both normalised by the wavelet validation loss at the first recorded training step. Lower panels show the low-pass MSE component and individual wavelet-band contributions as fractions of the total regional wavelet validation loss. Here, $j=1$ denotes the finest wavelet-band, with increasing $j$ corresponding to progressively coarser scales. All decomposition levels use uniform scale weights, $\alpha_j=1$. 
    }
    \label{fig:scale_loss_evolution}
\end{figure}

Having demonstrated the scale selectivity of the loss using controlled perturbations, we next examine how its scale-wise composition evolves during model training. Figure~\ref{fig:scale_loss_evolution} shows the magnitude and scale composition of the regional wavelet validation loss during the early adaptation of the model to the regional domain for three variables. These variables were selected to span a range of characteristic spatial structures represented within kilometre-scale atmospheric fields: \SI{500}{hPa} geopotential is dominated by smooth, large-scale variability; \SI{2}{m} temperature contains additional structure associated with orography, surface heterogeneity, and land--sea contrasts; and precipitation is intermittent and contains substantial variability across intermediate and fine scales. The figure covers the first 25\% of the regional single-step training stage (Stage~3a; see Appendix~\ref{app:model_configuration} for the complete training schedule).

For \SI{500}{hPa} geopotential, the low-pass MSE contributes more than 90\% of the objective throughout the displayed period, and the pointwise MSE and complete wavelet loss (Eq.~\ref{eq:wavelet_total_loss}) follow similar trajectories. Although the absolute losses decrease at every wavelet level, the low-pass term remains dominant, consistent with the concentration of geopotential variance at large spatial scales.

For \SI{2}{m} temperature and precipitation, the absolute low-pass and wavelet-band losses also decrease, but the low-pass error falls more rapidly. The wavelet-band levels therefore account for an increasing fraction of the remaining wavelet objective. For temperature, the continued separation between the wavelet validation loss and pointwise MSE indicates residual scale-dependent errors that are not distinguished by the pointwise metric. These contributions are consistent with the greater regional and fine-scale variability of near-surface temperature. Precipitation contains substantial scale-dependent error from the beginning of training. Its pointwise MSE flattens comparatively early, whereas the wavelet validation loss continues to decrease, indicating continued improvement in aspects of the precipitation field represented by the multiscale objective. 

The three examples demonstrate that the distribution of the wavelet loss reflects the characteristic spatial scales of each variable. Broad-scale fields remain dominated by the low-pass component, whereas variables containing greater regional and fine-scale variability receive progressively larger contributions from the wavelet bands. The increasing relative wavelet-band contributions reflect differences in the rates at which errors are reduced, rather than growth in scale-dependent error. These diagnostics characterise the model's early adaptation to the regional domain and should not be interpreted as the complete evolution of the loss over the full training schedule.

\section{Forecast evaluation} 
\label{sec:forecast_evaluation} 

We evaluate three variants of a developmental joint global--regional configuration for a regional FastNet~\cite{daub2025,dunstan2026} model, implemented using the Anemoi~\cite{lang2024} framework. The variants share the same architecture and differ principally in their training objectives: pointwise MSE, an FFT-based variant of the AMSE loss introduced by Subich et al.~\cite{subich2025}, and the wavelet loss introduced here. For the wavelet-trained model, all decomposition levels are weighted uniformly, with $\alpha_j=1$. Each model produces forecasts on a global N320 reduced Gaussian grid ($\sim\SI{31}{km}$) and a \SI{1.5}{km} regional grid centred on the UK, using a separate encoder--decoder pair for each dataset. Verifying forecasts are initialised every \SI{6}{hours}, with output at \SI{6}{hour} intervals to a lead time of \SI{240}{hours}, and are produced for the entirety of 2023. Full details of the model configuration and the training schedule are provided in Appendix~\ref{app:model_configuration}. 

Forecasts are verified against surface observations using pointwise RMSE and the High-Resolution Assessment (HiRA) framework~\cite{Mittermaier2014}. HiRA treats the model grid points surrounding each observation as a pseudo-ensemble, which we evaluate using the continuous ranked probability score (CRPS). This neighbourhood-based approach provides greater tolerance to small spatial displacement errors than pointwise verification and assesses whether forecasts contain locally relevant structure. We calculate HiRA CRPS using neighbourhoods of $3\times3$, $7\times7$, and $11\times11$ model grid points. Details of the observation processing and forecast--observation matching are provided in Appendix~\ref{app:verification_methodology}, with additional surface and upper-air verification presented in Appendix~\ref{app:additional_verification}.

\section{Results}

\subsection{Verification against surface observations}

\begin{figure}[ht]
    \centering
    \includegraphics[width=\linewidth]{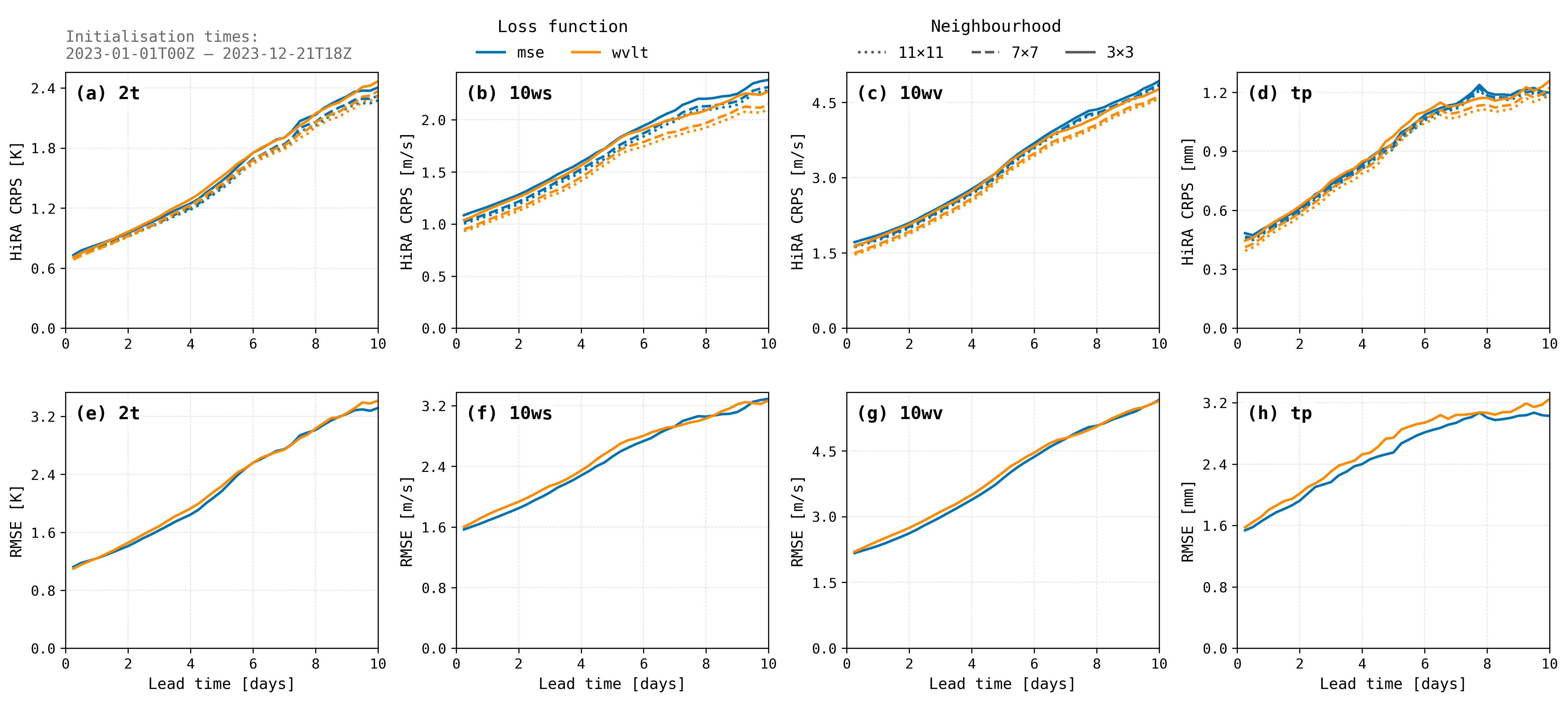}
    \caption{
    Verification of the MSE- and wavelet-trained models against surface observations as a function of lead time. The top row shows HiRA CRPS for $3\times3$, $7\times7$ and $11\times11$ model-grid neighbourhoods; the bottom row shows pointwise RMSE. Columns show (a,e) \SI{2}{m} temperature, (b,f) \SI{10}{m} wind speed, (c,g) \SI{10}{m} wind-vector error and (d,h) \SI{6}{hour} accumulated precipitation. Curves are aggregated over forecasts initialised between 1 January and 21 December 2023. Lower values indicate better performance.
    }
    \label{fig:obs_verification}
\end{figure}

The models show similar overall skill but their relative performance depends consistently on the verification metric (Figure~\ref{fig:obs_verification}). The wavelet-trained model generally achieves lower HiRA CRPS, with the clearest improvements for both \SI{10}{m} wind diagnostics across all three neighbourhood sizes. Smaller improvements are present for \SI{2}{m} temperature and precipitation over parts of the forecast, although the differences become less systematic at longer leads. In contrast, the MSE-trained model generally has slightly lower pointwise RMSE, particularly for precipitation and, to a lesser extent, temperature and wind diagnostics.

This contrast is consistent with the wavelet-trained model retaining more local spatial variability without uniformly improving the exact placement of forecast features. Retaining this structure can improve neighbourhood-based scores while remaining vulnerable to the double penalty imposed by pointwise RMSE. However, lower HiRA CRPS may also result from improved local bias, intensity or variability, and should not be interpreted solely as evidence of displaced but otherwise accurate forecast features.

\subsection{Spectral fidelity}
\label{sec:spectral_evaluation}

\begin{figure}[t]
    \centering
    \includegraphics[width=\linewidth]{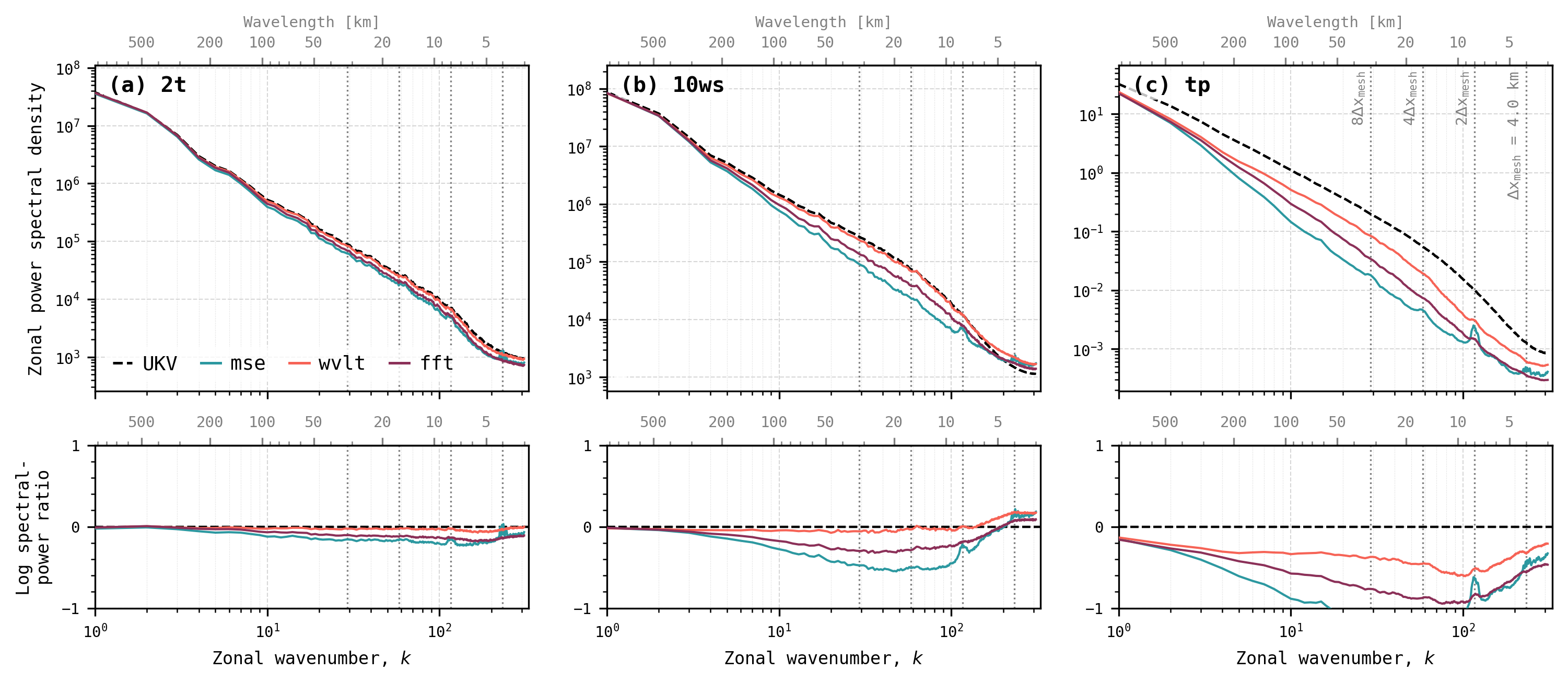}
    \caption{ 
    Zonal power spectra of forecasts from the MSE- (blue), FFT-AMSE- (purple), and wavelet-trained (orange) models at a lead time of 24 hours, verified against UKV operational analyses (black, dashed). Columns show (a) \SI{2}{m} temperature, (b) \SI{10}{m} wind speed, and (c) \SI{6}{hour} accumulated precipitation. Upper panels show the absolute spectra, while lower panels show $\log_{10}(P_f/P_r)$, where $P_f$ and $P_r$ are the forecast and reference powers, respectively. Zero indicates equal forecast and reference power, while negative values indicate a forecast power deficit. Vertical dotted lines mark the internal processor-mesh spacing, $\Delta x_{\mathrm{mesh}}=\SI{4}{km}$, and wavelengths corresponding to $2\Delta x_{\mathrm{mesh}}$, $4\Delta x_{\mathrm{mesh}}$, and $8\Delta x_{\mathrm{mesh}}$. 
    }
    \label{fig:power_spectra}
\end{figure}

The spectral differences among the models are substantially larger than their differences in pointwise and neighbourhood-based verification scores, illustrating the value of training objectives that explicitly constrain multiscale structure. Figure~\ref{fig:power_spectra} compares the three training objectives at a lead time of 24 hours; similar results are observed across lead times and are provided in Appendix~\ref{app:additional_verification} for reference. 

The MSE-trained model develops a systematic power deficit towards shorter wavelengths, consistent with the smoothing associated with pointwise optimisation. This deficit is modest for \SI{2}{m} temperature but pronounced for \SI{10}{m} wind speed and accumulated precipitation. FFT-AMSE recovers some of the variance lost under MSE training, but the wavelet-trained forecasts remain closer to the UKV reference over a broader range of intermediate and short wavelengths. The distinction is clearest for wind speed and precipitation, for which wavelet training produces a more accurate distribution of variance across the resolved scales. 

The MSE-trained spectra also contain regions of excess power aligned with wavelengths associated with the internal processor mesh, most clearly for precipitation. These features are absent or strongly suppressed in the wavelet-trained spectra, indicating that the improvement is not simply an indiscriminate increase in short-wavelength variance. Instead, the wavelet objective both reduces spectral power deficits and limits non-physical mesh-related artefacts~\cite{dunstan2026,pasquini2026}. 

\subsection{Precipitation forecasts}

\begin{figure}[ht]
    \centering
    \includegraphics[width=\linewidth]{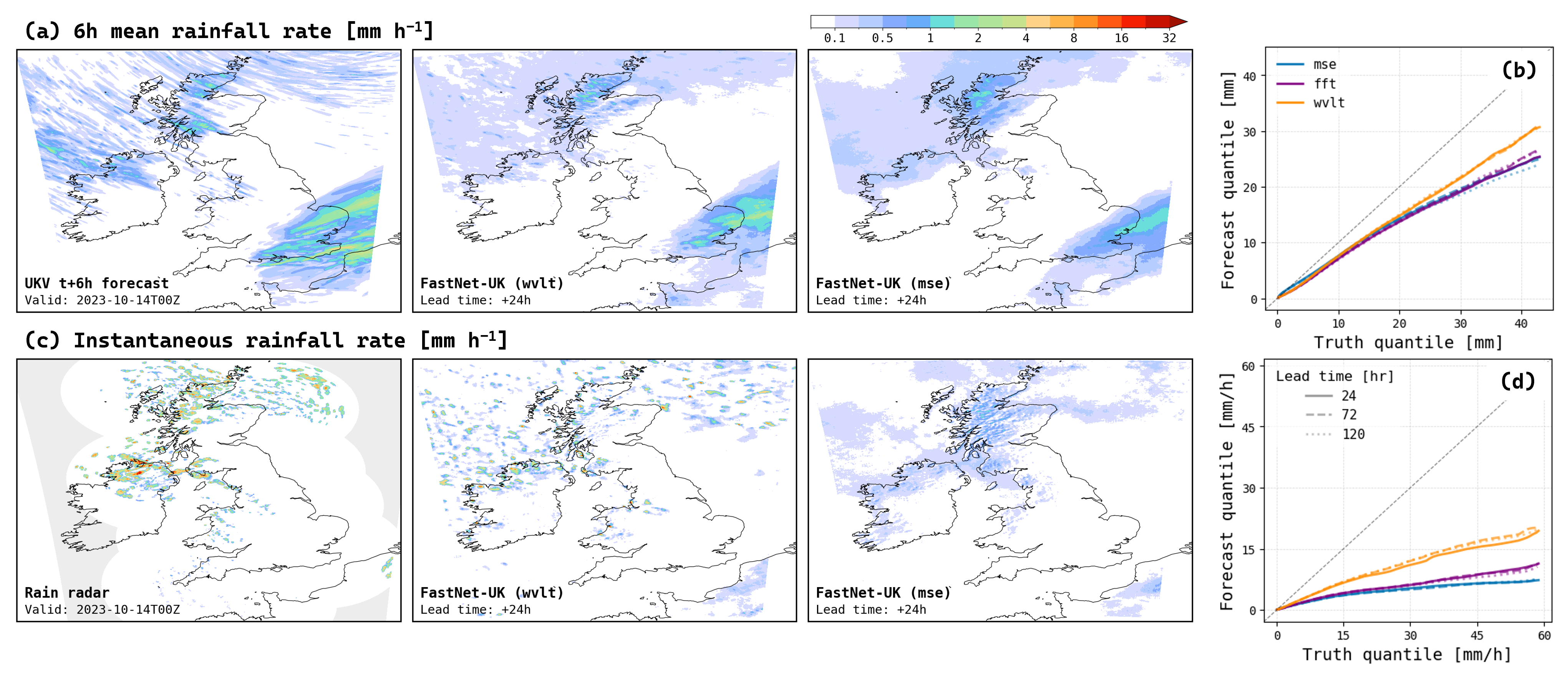}
    \caption{Precipitation forecast characteristics. (a) Mean rainfall rate over the preceding \SI{6}{hours} for the case valid at 0000 UTC on 14 October 2023, calculated by dividing the \SI{6}{hour} precipitation accumulations by six. The panels show the UKV operational NWP forecast at a lead time of \SI{6}{hours} and the wavelet- and MSE-trained model forecasts at a lead time of \SI{24}{hours}. (b) Quantile--quantile comparison of forecast and analysis \SI{6}{hour} accumulated precipitation over the 2023 verification period, comparing the MSE- (blue), FFT-AMSE- (purple), and wavelet-trained (orange) models at lead times of 24 (solid), 72 (dashed), and 120 (dotted) hours. (c) Instantaneous rainfall rate for the same case and valid time, with the model forecasts at a lead time of 24 hours verified against radar-derived rainfall estimates. Panels (a) and (c) use the same colour scale in \si{mm.h^{-1}}, such that a given colour represents the same rainfall rate in both rows. (d) As in (b), but for instantaneous rainfall rate verified against radar-derived rainfall estimates. In (b) and (d), the grey dashed diagonal indicates perfect agreement.}
    \label{fig:precip}
\end{figure}

Precipitation provides a stringent test of km-scale MLWP models because forecast quality depends on representing both the intensity and spatial organisation of highly intermittent small-scale features. The improved spectral fidelity obtained with wavelet training is reflected in both the structure and distribution of forecast precipitation (Figure~\ref{fig:precip}). 

The wavelet-trained model produces precipitation fields that are sharper and more intense than those generated by the MSE-trained model (Figure~\ref{fig:precip}a,c). The difference is particularly evident for instantaneous rainfall rates, where the MSE forecast is dominated by diffuse weak precipitation, while the wavelet-trained forecast contains more coherent and spatially concentrated features that more closely resemble the radar observations. Similar behaviour is evident for accumulated precipitation, with stronger local maxima and sharper gradients. 

These qualitative differences are reflected in the precipitation distributions. For both accumulated precipitation and instantaneous rainfall rate, the wavelet-trained model more closely reproduces the upper tail of the observed distribution than either the MSE- or FFT-AMSE-trained models (Figure~\ref{fig:precip}b,d). FFT-AMSE partially reduces the underprediction of heavy precipitation relative to MSE, while wavelet training provides a further improvement. This greater intensity and fine-scale variability is accompanied by improved spatial skill: at a lead time of \SI{24}{hours}, the wavelet-trained model achieves higher Fractions Skill Scores than both alternatives across most precipitation thresholds and neighbourhood sizes, with the clearest gains for heavier precipitation and broader neighbourhoods (Appendix~\ref{app:fss}). These combined improvements are consistent with benefits from the spatially localised wavelet representation beyond those obtained using Fourier-based spectral constraints alone.

Despite these improvements, precipitation remains a clear weakness of all model variants. The highest quantiles are systematically underestimated, particularly for instantaneous rainfall rate, and substantial departures from one-to-one agreement persist even at short lead times. Furthermore, although wavelet training produces sharper accumulated precipitation fields, neither MLWP model fully reproduces the elongated streak-like structures evident in the operational forecast that result from precipitation features advecting during the six-hour accumulation period. The wavelet objective therefore improves the representation of heavy precipitation but does not eliminate important shortcomings in the temporal organisation, morphology and intensity of precipitation.

\section{Conclusions} 
\label{sec:conclusions} 

We have introduced a deterministic training objective for machine-learning weather prediction based on the two-dimensional dual-tree complex wavelet transform. The DT$\mathbb{C}$WT decomposes atmospheric fields jointly by scale, location, and orientation, allowing amplitude- and coherence-sensitive loss terms to be evaluated independently at each spatial scale. Its compactly supported basis functions also produce a more spatially localised training gradient than an equivalent Fourier-domain objective, while its approximate shift invariance and directional selectivity are well suited to sharply localised atmospheric structures. 

Applied to an MLWP model that simultaneously produces \SI{1.5}{km} forecasts over the UK and $\sim\SI{31}{km}$ forecasts globally, the wavelet loss provides a better balance between spatial accuracy and multiscale fidelity than either pointwise MSE or FFT-AMSE. Within the regional domain, the wavelet-trained model retains pointwise performance close to the MSE baseline, generally improves neighbourhood-based verification, and outperforms FFT-AMSE. The distinction is clearest in spectral verification: across the evaluated variables and lead times, the wavelet-trained forecasts remain closer to the reference spectra over a broader range of resolved scales than either baseline. Wavelet training substantially reduces the intermediate- and small-scale power deficits produced by MSE and suppresses non-physical excess power aligned with the internal processor-mesh scales. Compared with FFT-AMSE, it more accurately reproduces the reference distribution of variance across spatial scales without incurring the same degradation in pointwise performance.

These improvements are reflected in sharper, more spatially concentrated precipitation forecasts and more realistic intensity distributions than those produced by either baseline. Nevertheless, all three models underestimate the most intense rainfall, and wavelet training does not fully recover the morphology and temporal organisation of accumulated precipitation. The wavelet loss therefore reduces, but does not eliminate, the trade-off between preserving fine-scale variability and maintaining spatial accuracy. Future work should assess its sensitivity to scale weighting, evaluate its generality across additional regions and weather regimes, and investigate its use within probabilistic forecasting systems that represent both fine-scale structure and forecast uncertainty.

\section*{Acknowledgments}
The development of FastNet, including this work and its contributors, is funded by The Alan Turing Institute and the Met Office AWAKEN project. The AWAKEN project is led by the UK Met Office and is funded through the UK Government's International Science Partnerships Fund (ISPF), with support from the Department for Business, Innovation, Science and Trade (DBIST) and the Foreign, Commonwealth and Development Office (FCDO). The authors acknowledge the use of resources provided by the Isambard-AI National AI Research Resource (AIRR). Isambard-AI is operated by the University of Bristol and is funded by the UK Government’s Department for Business, Innovation, Science and Trade (DBIST) via UK Research and Innovation; and the Science and Technology Facilities Council [ST/AIRR/I-A-I/1023]. The authors gratefully acknowledge all contributors to the
Anemoi framework.

\subsection*{Statement on the use of generative AI}
All results, analysis and scientific conclusions in this manuscript were produced by the authors. Microsoft 365 Copilot was used for editorial assistance to improve clarity and readability.

\printbibliography

\newpage

\appendix
\renewcommand{\thefigure}{\Alph{section}\arabic{figure}} 
\counterwithin{figure}{section}
\renewcommand{\thetable}{\Alph{section}\arabic{table}} 
\counterwithin{table}{section} 
\renewcommand{\theequation}{\Alph{section}\arabic{equation}} 
\counterwithin{equation}{section}

\section{Spatial attribution and gradient comparison} \label{app:spatial_attribution} 

\begin{figure}[!h] 
\centering 
\includegraphics[width=1\linewidth]{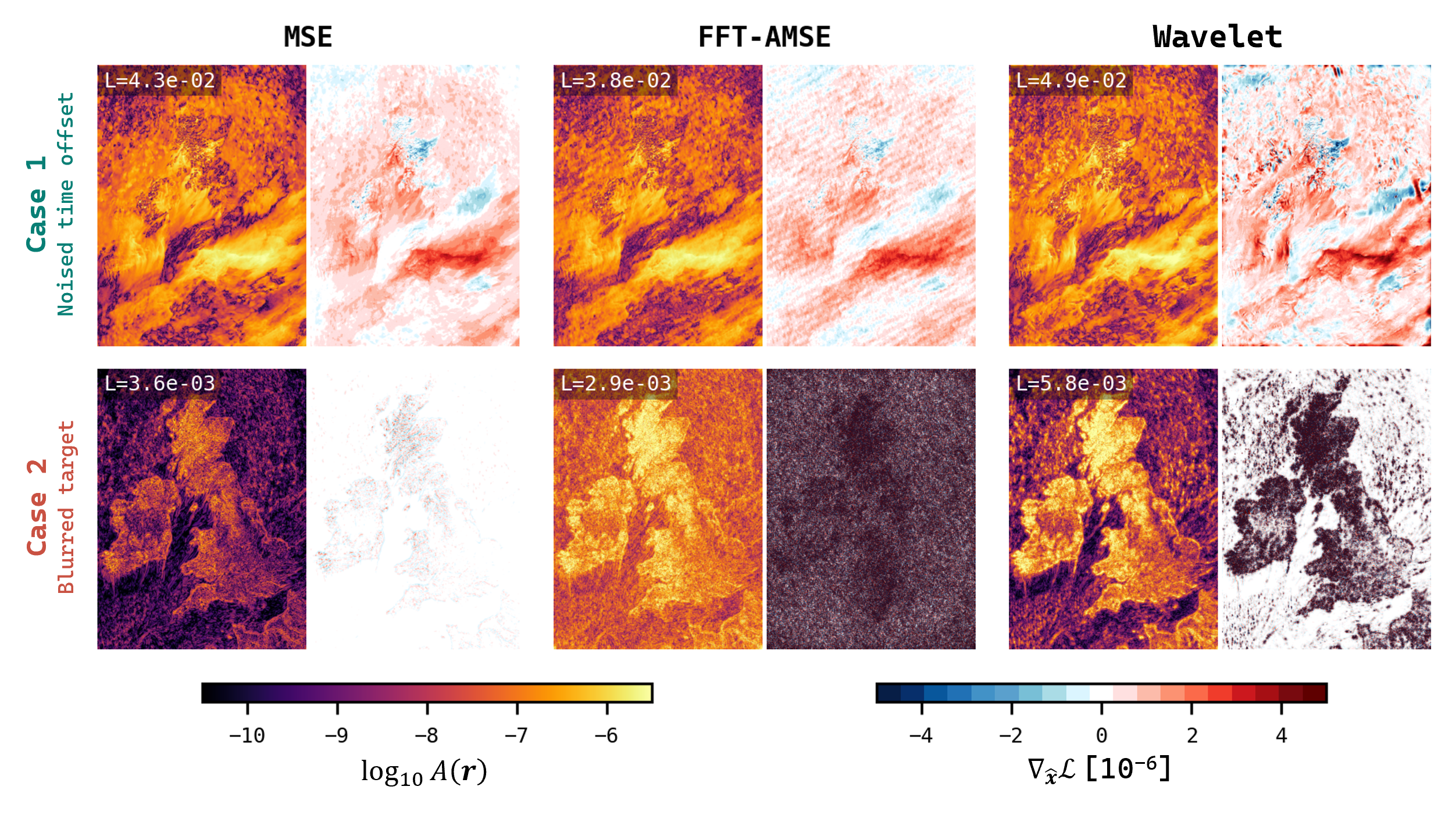} 
\caption{Comparison of first-order spatial attribution and field-space gradients for \SI{2}{m} temperature using MSE, Fourier-domain AMSE (FFT-AMSE), and the wavelet-based loss. Rows show the two constructed prediction fields introduced in Figure~\ref{fig:wavelet_decomposition}: a perturbed \SI{6}{hour} time-offset prediction with added Gaussian noise (Case 1), and a Gaussian-blurred target with $\sigma=3$ grid cells (Case 2). For each objective, the left panel shows the attribution $A(\boldsymbol{r})$ defined by Equation~\ref{eq:spatial_attribution}, and the right panel shows the field-space gradient $\partial\mathcal{L}/\partial\widehat{x}(\boldsymbol{r})$. Attribution is displayed on a logarithmic scale. Common colour scales are used across objectives for each diagnostic. Attribution combines the local prediction error with the sensitivity of the objective and is therefore a first-order diagnostic rather than an exact spatial decomposition of the scalar objective. Absolute loss and gradient magnitudes depend on the definition and normalisation of each objective, so the comparison emphasises the spatial structure of the resulting fields. } 
\label{fig:loss_comparison} 
\end{figure} 

Section~\ref{sec:scale_selectivity} showed how the wavelet objective distributes its scalar loss and field-space training signal among decomposition levels. Here, we compare the spatial structure of the complete training signals produced by pointwise MSE, FFT-AMSE, and the wavelet-based loss for the same two constructed prediction fields. 

For a scalar objective $\mathcal{L}$, prediction $\widehat{\boldsymbol{x}}$, and target $\boldsymbol{x}$, we define the first-order spatial attribution at location $\boldsymbol{r}$ as 

\begin{equation} 
\begin{aligned} 
A(\boldsymbol{r}) = \left| \left[ \widehat{x}(\boldsymbol{r}) - x(\boldsymbol{r}) \right] \frac{ \partial\mathcal{L} }{ \partial\widehat{x}(\boldsymbol{r}) } \right|. 
\end{aligned} 
\label{eq:spatial_attribution} 
\end{equation} 

This diagnostic combines the magnitude of the local prediction error with the sensitivity of the complete objective at the same location. It represents the magnitude of the first-order change associated with moving that prediction value towards its target. For pointwise MSE, the attribution is proportional to the pointwise squared error. For transform-domain objectives such as FFT-AMSE and the wavelet-based loss, coefficients from multiple spatial locations are aggregated and coupled within the objective. The attribution is therefore not an exact additive decomposition of the finite scalar loss, but provides a common field-space diagnostic of where the different objectives are most sensitive. 

Figure~\ref{fig:loss_comparison} shows the attribution and field-space gradient for each objective and prediction regime. Attribution is shown logarithmically because it spans several orders of magnitude. The gradient is displayed separately because the two diagnostics answer different questions: the gradient shows the signed sensitivity of the objective to changes in the predicted field, whereas the attribution also weights that sensitivity by the prediction error already present. A location can therefore have a large gradient but modest attribution when its error is small, or large attribution when both error and sensitivity are appreciable. 

For the perturbed time-offset prediction (Case 1), all three objectives produce broadly similar attribution patterns, highlighting many of the same broad and finer-scale errors. Their field-space gradients nevertheless differ. The MSE gradient follows the pointwise prediction error, consistent with the local definition of the objective. The FFT-AMSE gradient is more broadly distributed across the domain, consistent with the global support of the Fourier basis. The wavelet gradient retains stronger location-dependent variation around the atmospheric structures contributing to the error. This behaviour is consistent with the spatial indexing and compact support of the wavelet coefficients, despite the coupling introduced by aggregation within each scale--orientation band. 

The contrast is stronger for the blurred prediction (Case 2). The MSE gradient follows the pointwise residual introduced by smoothing but does not distinguish the spatial scales contributing to that residual. FFT-AMSE responds to the missing spectral variance, but its gradient is distributed throughout much of the domain. The wavelet objective is similarly sensitive to the removed variability, while producing a gradient that is more strongly organised around the temperature features altered by smoothing. The response is concentrated primarily over land and is particularly structured over regions of complex orography, including the Scottish Highlands, with a weaker response over much of the surrounding sea. This organisation is consistent with fine-scale \SI{2}{m} temperature variability associated with terrain and land--sea contrasts. 

Together with the scale-wise decomposition in Figure~\ref{fig:wavelet_decomposition}, this comparison shows that the wavelet objective both identifies the affected range of spatial scales and retains location-dependent sensitivity to the corresponding atmospheric structures. Absolute loss and gradient magnitudes depend on the definition and normalisation of each objective and are not directly comparable; the comparison is therefore restricted to the spatial structure and relative localisation of the attribution and gradient fields.

\section{Model configuration and training schedule} 
\label{app:model_configuration} 

The model was developed and trained using the Anemoi framework~\cite{lang2024}, and follows an encode--process--decode architecture with graph-transformer GNNs used in all three components. Input fields are encoded onto a latent mesh, processed in latent space and decoded onto the output grids. Separate encoder--decoder pairs are used for two datasets: ERA5 on its native reduced Gaussian grid, representing the global atmospheric state, and Met Office UKV operational analyses on a regional grid covering the UK at approximately \SI{1.5}{km} resolution. This configuration allows the datasets to use different grids and variables while sharing a common latent representation. The model therefore produces global and high-resolution regional forecasts simultaneously in a single forward pass. 

The MSE-, FFT-AMSE- and wavelet-trained models use the staged training schedule summarised in Table~\ref{tab:fastnet_uk_training}, following the approach introduced in Ref.~\cite{2026nipen}. Stage~1 establishes the global forecasting capability using ERA5 on the O96 grid and a coarse global latent mesh. In Stage~2, the resolution of the ERA5 dataset is increased to N320 and the global latent mesh is refined, allowing the model to learn a higher-resolution global representation over the same training period. 

Stage~3a introduces the UKV dataset, together with a regional encoder--decoder pair and a high-resolution regional mesh embedded within the global mesh. The shared latent representation retains the global atmospheric context while adapting the model to the UK domain. Global and regional losses are assigned relative weights of 67\% and 33\%, respectively. Without this weighting, the regional domain would contribute less than 1\% of the combined objective because of its much smaller spatial extent. 

Stage~3b extends training from single-step forecasts to autoregressive rollouts of 2--12 steps, exposing the model to its own predictions and improving stability at longer lead times. The reduced learning rate in this stage allows the model to be fine-tuned over longer forecast horizons while limiting degradation of the global and regional representations learned in the preceding stages. 

\begin{table}[t]
\centering
\small
\renewcommand{\arraystretch}{1.1}
\setlength{\tabcolsep}{6pt}
\captionsetup{skip=3pt}

\begin{threeparttable}
\caption{Training configuration for each stage of the model used in this study. A dash indicates that no regional dataset or regional mesh was used in that stage. All training was performed on the Isambard-AI (Phase 2) HPC system~\cite{Isambard-AI} using NVIDIA GH200 superchips.}
\label{tab:fastnet_uk_training}
\begin{tabular*}{\textwidth}{@{\extracolsep{\fill}} l c c c c}
\toprule
& \multirow{2}{*}{\textbf{Stage 1}}
& \multirow{2}{*}{\textbf{Stage 2}}
& \multicolumn{2}{c}{\textbf{Stage 3}} \\
\cmidrule(lr){4-5}
& & & \textbf{a} & \textbf{b} \\
\midrule

\textbf{Global dataset} & ERA5 & ERA5 & ERA5 & ERA5 \\
\addlinespace[0.3em]

\textbf{Global data grid} & O96 & N320 & N320 & N320 \\
\addlinespace[0.3em]

\textbf{Global resolution} & 100 km & 30 km & 30 km & 30 km \\
\addlinespace[0.3em]

\textbf{Global mesh refinement} & 5 & 6 & 6 & 6 \\
\addlinespace[0.3em]

\textbf{Global wavelet levels, $J$} & 4 & 6 & 6 & 6 \\
\addlinespace[0.3em]

\textbf{Regional dataset} & -- & -- & UKV & UKV \\
\addlinespace[0.3em]

\textbf{Regional resolution} & -- & -- & 1.5 km & 1.5 km \\
\addlinespace[0.3em]

\textbf{Regional mesh refinement} & -- & -- & 11 & 11 \\
\addlinespace[0.3em]

\textbf{Regional wavelet levels, $J$} & -- & -- & 6 & 6 \\
\addlinespace[0.3em]

\textbf{Rollout} & 1 & 1 & 1 & 2--12 \\
\addlinespace[0.3em]

\textbf{Training period} & 1980--2022 & 1980--2022 & 2020--2022 & 2020--2022 \\
\addlinespace[0.3em]

\textbf{Iterations} & 100,000 & 50,000 & 25,000 & 3200 \\
\addlinespace[0.3em]

\textbf{Peak learning rate} & $2\times10^{-3}$ & $3.2\times10^{-4}$ & $4\times10^{-4}$ & $2.5\times10^{-5}$ \\
\addlinespace[0.3em]

\textbf{Warm-up iterations} & 1000 & 100 & 500 & 50 \\
\addlinespace[0.3em]

\textbf{Batch size} & 128 & 32 & 32 & 32 \\
\addlinespace[0.3em]

\textbf{Total number of GPUs} & 16 & 32 & 64 & 64 \\
\addlinespace[0.3em]

\textbf{GPUs per model instance} & 1 & 1 & 2 & 2 \\
\addlinespace[0.3em]

\textbf{Total GPU hours} & 540 & 720 & 800 & 400 \\

\bottomrule
\end{tabular*}

\begin{tablenotes} 
\footnotesize 
\item[] 
\textit{Note.} Mesh refinement denotes the subdivision frequency of the icosahedron used to construct the latent mesh with larger values corresponding to finer meshes. Refinement frequencies of 6 and 11 give average node spacings of approximately \SI{110}{km} and \SI{4}{km}, respectively. Wavelet-transform settings apply only to the wavelet-trained experiment; all other configuration settings are shared by the three model variants. 
\end{tablenotes}

\end{threeparttable}
\end{table}

The three models differ only in their training objectives. The MSE- and wavelet-loss trained models use their respective loss functions throughout all four stages and for both the global and regional components. The FFT-AMSE model used a fast Fourier Transform (FFT) variant of the AMSE loss introduced in Ref.~\cite{subich2025} in the regional component, replacing the spherical harmonic transform with an FFT performed over the regional domain. Its global component uses MSE and is initialised from the MSE model at the end of Stage 2.

For the wavelet experiment, fields on reduced Gaussian grids are regridded to regular latitude--longitude grids before applying the two-dimensional DT$\mathbb{C}$WT: O96 fields are mapped to a \SI{1.0}{\degree} grid and N320 fields to a \SI{0.25}{\degree} grid, following Ref.~\cite{dunstan2026}. UKV fields are transformed directly on the native regional grid. The number of wavelet decomposition levels used for each dataset is given in Table~\ref{tab:fastnet_uk_training}, and is calculated via

\begin{equation} 
J = \left\lfloor \log_2\left[\min(H,W)\right] \right\rfloor-3, 
\label{eq:number_wavelet_levels} 
\end{equation} 

where $H$ and $W$ are the meridional and zonal dimensions of the input grid, respectively.

We compute the dual-tree complex wavelet transform using a custom Triton~\cite{triton} implementation written by the authors, validating its forward transform and gradients against \texttt{pytorch\_wavelets}~\cite{pytorch_wavelets,cotter_thesis_2019} to within numerical precision. The implementation uses the \texttt{`near\_sym\_b'} and \texttt{`qshift\_b'} filter sets provided by \texttt{pytorch\_wavelets}, corresponding to a first-level biorthogonal filter bank followed by quarter-shift filters at subsequent levels~\cite{selesnick2005}. With the Triton implementation, wavelet training achieved approximately 20\% lower throughput than the MSE baseline in Stages~1 and~2. No difference was observed in Stage~3, where throughput was limited by data loading rather than loss computation.

\section{Verification methodology} 
\label{app:verification_methodology} 

Surface verification uses quality-controlled observations from 114 stations across the UK. Six-hour precipitation accumulations are derived separately from hourly Synoptic Rainfall Europe West observations, following standard operational verification procedures. The same observation datasets and processing are used to evaluate all model configurations.

Neighbourhood-based verification is performed using the High-Resolution Assessment (HiRA) framework~\cite{Mittermaier2014}. For each observation, a square neighbourhood centred on the nearest model grid point is interpreted as a pseudo-ensemble, and the resulting distribution is evaluated using the continuous ranked probability score (CRPS). We use neighbourhoods of $3\times3$, $7\times7$, and $11\times11$ model grid points. HiRA therefore provides some tolerance to small spatial displacement errors while still rewarding appropriate local values and variability. 

For variables without suitable station observations, forecasts are verified against the corresponding UKV operational analyses. For each analysis field, 1,000 grid points are sampled at fixed, reproducibly selected locations within the regional domain and treated as pseudo-observations. At each location, the surrounding forecast grid points form the HiRA pseudo-ensemble, from which CRPS is calculated for neighbourhoods of $3\times3$, $7\times7$, and $11\times11$ model grid points. These scores measure agreement with the gridded analysis rather than independent observational skill.

Pointwise RMSE is calculated using variable-specific collocation. Forecast \SI{2}{m} temperature is bilinearly interpolated to each observation location, and observations are adjusted to the UKV model orography using a standard environmental lapse rate of \SI{-0.0065}{K\,m^{-1}}~\cite{wallace2006}. The same elevation-adjusted observations are used for the corresponding HiRA calculation. Six-hour precipitation accumulations are compared with the nearest model grid point to avoid smoothing intermittent precipitation fields. The \SI{10}{m} zonal and meridional wind components are bilinearly interpolated before calculating wind speed and vector error.

\section{Additional verification} 
\label{app:additional_verification} 

\subsection{Comparison with the MSE-trained model} 
\label{app:mse_wavelet_comparison} 

\begin{figure}[ht] 
\centering 
\includegraphics[width=0.8\linewidth]{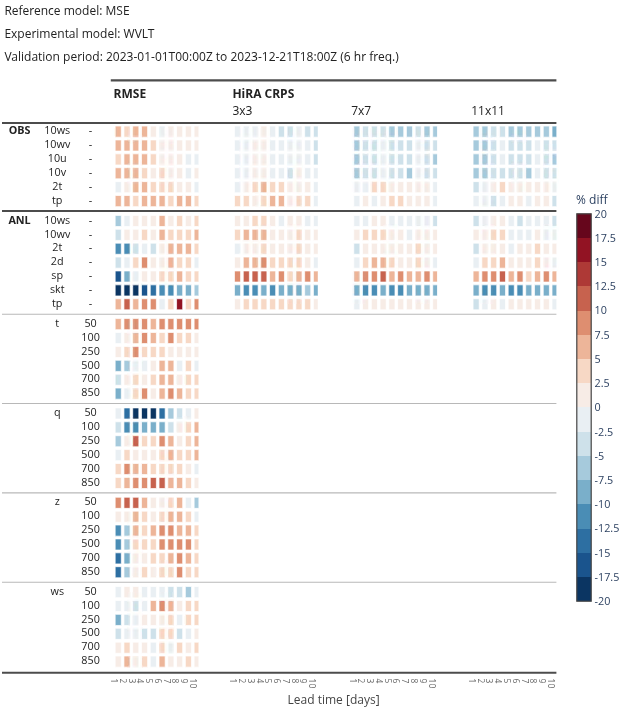} 
\caption{ Scorecard showing the percentage difference in verification scores between the wavelet- and MSE-trained models over lead times of up to 10 days. ``OBS'' and ``ANL'' denote verification against observations and operational analyses, respectively. Surface variables are \SI{10}{m} wind speed (10ws), \SI{10}{m} wind vector error (10wv), the zonal and meridional components of \SI{10}{m} wind (10u and 10v), \SI{2}{m} temperature (2t), total precipitation (tp), surface pressure (sp) and skin temperature (skt). Upper-air variables are temperature (t), specific humidity (q), geopotential (z), and wind speed (ws), with the corresponding pressure levels given in \si{hPa}. Columns show pointwise root-mean-squared error (RMSE) and high-resolution assessment (HiRA) continuous ranked probability score (CRPS), with HiRA CRPS calculated using neighbourhoods of $3\times3$, $7\times7$, and $11\times11$ model grid points. Scores are aggregated over forecasts initialised between 1 January and 21 December 2023. Blue indicates lower errors for the wavelet-trained model, whereas red indicates lower errors for the MSE-trained model. }
\label{fig:mse_vs_wvlt_scorecard} 
\end{figure} 

\begin{figure}[ht]
    \centering
    \includegraphics[width=\linewidth]{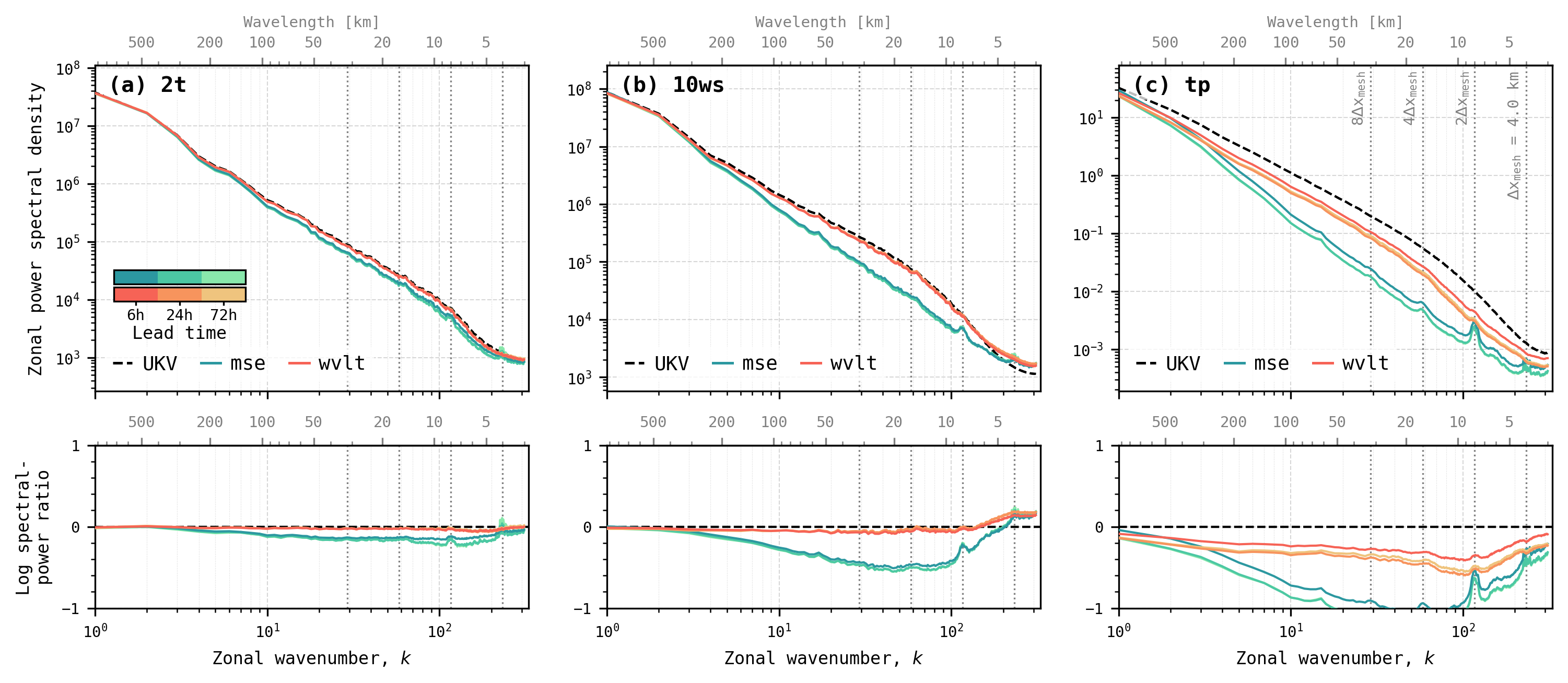}
    \caption{ 
    Lead-time dependence of the zonal power spectra for the MSE- (blue/green) and wavelet-trained (orange) models, verified against UKV operational analyses (black, dashed). Columns show (a) \SI{2}{m} temperature, (b) \SI{10}{m} wind speed, and (c) \SI{6}{hour} accumulated precipitation. Upper panels show the absolute spectra at lead times of 6, 24, and 72 hours, with colour shade indicating lead time. Lower panels show $\log_{10}(P_f/P_r)$, where $P_f$ and $P_r$ are the forecast and reference powers, respectively. Vertical dotted lines mark the internal processor-mesh spacing, $\Delta x_{\mathrm{mesh}}=\SI{4}{km}$, and wavelengths corresponding to $2\Delta x_{\mathrm{mesh}}$, $4\Delta x_{\mathrm{mesh}}$, and $8\Delta x_{\mathrm{mesh}}$. The reduction in intermediate- and short-wavelength power deficits under wavelet training persists across the evaluated lead times. 
    } 
    \label{fig:power_spectra_lead_time}
\end{figure}

\begin{figure}[ht] 
\centering 
\includegraphics[width=\linewidth]{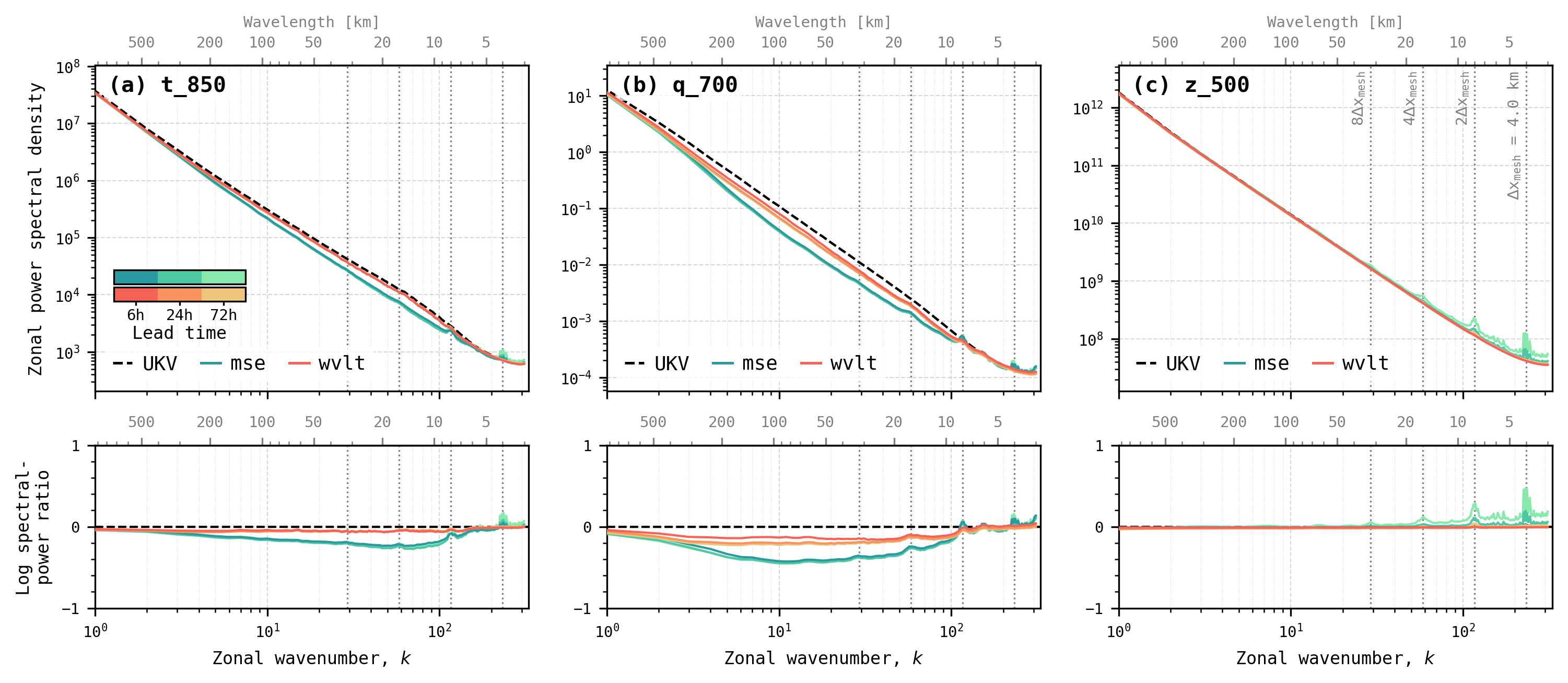} 
\caption{ 
As Figure~\ref{fig:power_spectra_lead_time}, but for upper-air variables: (a) \SI{850}{hPa} temperature, (b) \SI{700}{hPa} specific humidity, and (c) \SI{500}{hPa} geopotential. 
} 
\label{fig:upper_air_spectra} 
\end{figure}

Figure~\ref{fig:mse_vs_wvlt_scorecard} extends the comparison between the MSE- and wavelet-trained models across surface and upper-air variables, verification metrics, and lead times. For verification against surface observations, wavelet training generally improves HiRA CRPS, particularly for the wind diagnostics and at larger neighbourhood sizes, while the MSE-trained model generally retains a small advantage in pointwise RMSE. Differences in upper-air RMSE are also generally small and vary with variable and lead time.

The corresponding spectral differences are larger and more systematic. For the surface variables, the reduction in intermediate- and short-wavelength power deficits under wavelet training persists across lead times of 6, 24, and 72 hours (Figure~\ref{fig:power_spectra_lead_time}). The distinction is clearest for \SI{10}{m} wind speed and accumulated precipitation, while the differences for \SI{2}{m} temperature are smaller. This lead-time comparison supports the conclusions drawn from the 24-hour spectra in Section~\ref{sec:spectral_evaluation}, showing that the improved agreement with the reference spectra is not confined to a single forecast lead time. 

Similar behaviour is found for the upper-air variables (Figure~\ref{fig:upper_air_spectra}). The MSE-trained model develops intermediate-scale power deficits for \SI{850}{hPa} temperature and \SI{700}{hPa} specific humidity, both of which are substantially reduced by wavelet training. For \SI{500}{hPa} geopotential, the MSE-trained spectra instead contain narrow regions of excess power at wavelengths associated with the internal processor mesh; these features are strongly suppressed in the wavelet-trained model. 

Together, the additional verification shows that wavelet training produces forecasts that remain closer to the reference spectra across a broader range of resolved scales, variables, and lead times, while its effect on pointwise and neighbourhood-based scores depends more strongly on the variable and verification metric.

\subsection{Comparison with the FFT-AMSE-trained model} 
\label{app:fft_amse_wavelet_comparison} 

\begin{figure}[!t] 
\centering 
\includegraphics[width=0.8\linewidth]{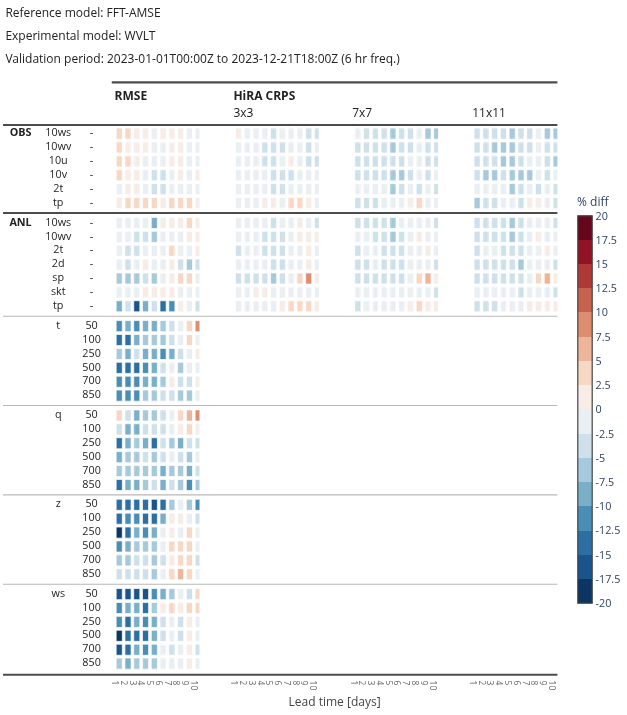} 
\caption{ Scorecard showing the percentage difference in verification scores between the wavelet- and FFT-AMSE-trained models over lead times of up to 10 days. ``OBS'' and ``ANL'' denote verification against observations and operational analyses, respectively. Columns show pointwise RMSE and HiRA CRPS calculated using neighbourhoods of $3\times3$, $7\times7$, and $11\times11$ model grid points. Scores are aggregated over forecasts initialised between 1 January and 21 December 2023. Blue indicates lower errors for the wavelet-trained model, whereas red indicates lower errors for the FFT-AMSE-trained model. } 
\label{fig:fft_amse_vs_wvlt_scorecard} 
\end{figure}

The wavelet-trained model generally outperforms the FFT-AMSE-trained model across variables, verification metrics, and lead times (Figure~\ref{fig:fft_amse_vs_wvlt_scorecard}). Improvements are particularly widespread for upper-air RMSE and are largest at short to medium lead times. Wavelet training also generally improves HiRA CRPS for surface variables, particularly at the larger neighbourhood sizes. Differences in surface RMSE are smaller and less systematic, although they predominantly favour the wavelet-trained model. 

Spectral fine-tuning commonly degrades pointwise performance relative to MSE because preserving fine-scale variability limits the smoothing that minimises pointwise error. The results presented here show that wavelet training substantially reduces this trade-off. The wavelet-trained model retains pointwise performance close to that of the MSE-trained model and performs considerably better than FFT-AMSE across most pointwise and neighbourhood-based metrics. At the same time, the spectral comparisons presented in Section~\ref{sec:spectral_evaluation} show that the wavelet-trained forecasts remain closer to the reference spectra over a broader range of resolved scales than either baseline. MSE training produces the largest intermediate- and small-scale power deficits, while FFT-AMSE recovers some of this missing variance but remains less consistent with the reference than the wavelet-trained model. Wavelet training therefore reproduces the reference distribution of variance across spatial scales more accurately than FFT-AMSE, while maintaining pointwise performance much closer to that of MSE. Among the three training objectives, the wavelet loss consequently provides the most favourable combination of forecast accuracy and physically realistic multiscale structure.

\subsection{Neighbourhood precipitation verification} 
\label{app:fss} 

\begin{figure}[!t] 
\centering 
\includegraphics[width=\linewidth]{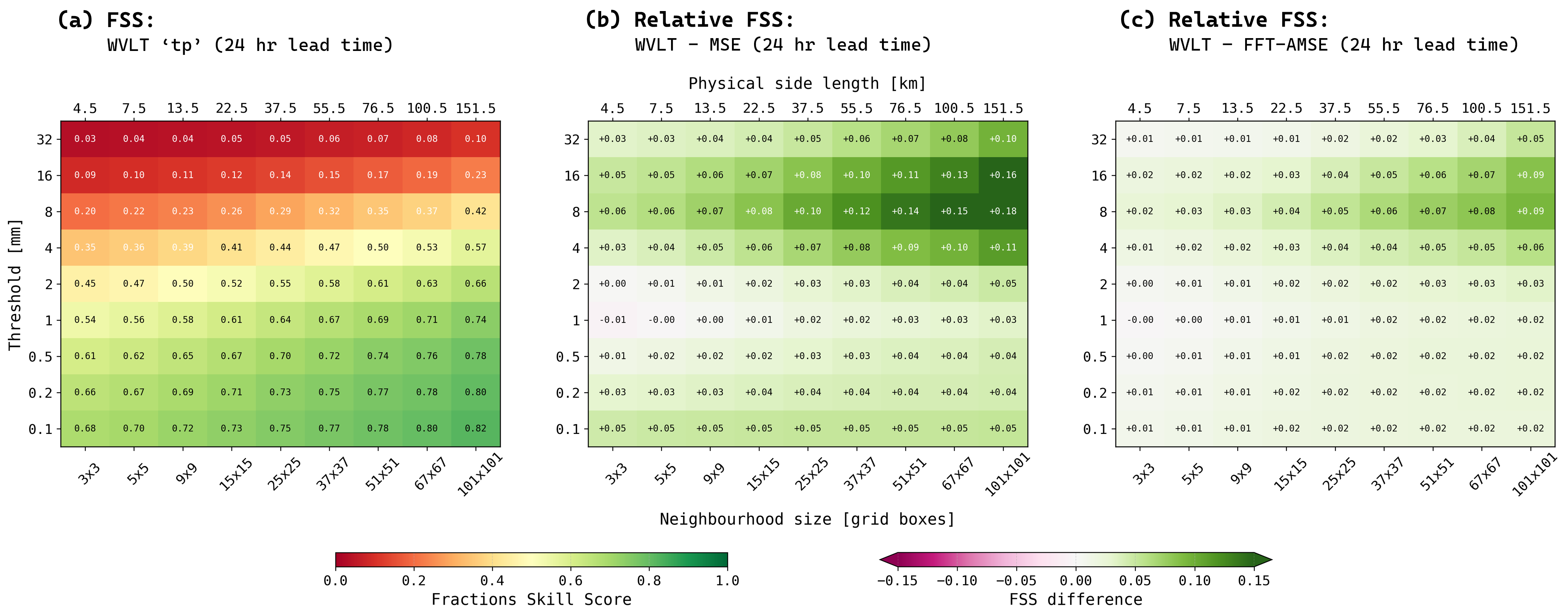} 
\caption{ Fractions Skill Score (FSS) for \SI{6}{hour} accumulated precipitation, `tp', at a lead time of \SI{24}{hours}, evaluated across precipitation thresholds and neighbourhood sizes over the regional domain. (a) Absolute FSS for the wavelet-trained model. (b,c) FSS difference between the wavelet-trained model and the MSE- and FFT-AMSE-trained models, respectively; positive values (green) indicate improved performance for the wavelet-trained model. Physical neighbourhood scales, calculated from the regional grid spacing, are shown along the upper axis. The same difference colour scale is used in panels (b) and (c). } 
\label{fig:fss_comparison} 
\end{figure}

The spectral and intensity-distribution diagnostics show that wavelet training produces sharper precipitation fields with greater fine-scale variability. Although HiRA CRPS assesses the continuous distribution of forecast values within a neighbourhood, it does not directly isolate performance for particular precipitation intensities. We therefore evaluate the Fractions Skill Score (FSS)~\cite{roberts2008} across a range of precipitation thresholds and neighbourhood sizes to determine how spatial skill varies with both event intensity and scale. FSS compares the fractions of forecast and observed grid points exceeding a given threshold within each neighbourhood. It ranges from zero to one, with one indicating perfect agreement, and provides a displacement-tolerant measure of precipitation skill that avoids the double penalty associated with strict pointwise verification. 

Figure~\ref{fig:fss_comparison} shows the absolute FSS of the wavelet-trained model and its differences relative to the MSE- and FFT-AMSE-trained models at a lead time of \SI{24}{hours}. Wavelet training improves FSS relative to MSE across almost all precipitation thresholds and neighbourhood sizes, with the largest gains occurring for heavier precipitation and broader neighbourhoods. Improvements over FFT-AMSE are smaller but remain consistently positive across most of the matrix and follow a similar dependence on threshold and neighbourhood size. 

The improvement over FFT-AMSE indicates that the benefit of wavelet training is not explained solely by greater spectral amplitude. The spatial localisation of the wavelet objective also improves the neighbourhood-scale agreement of precipitation structures. Wavelet training therefore produces sharper and more intense precipitation features without sacrificing spatial skill, with particularly clear benefits under displacement-tolerant evaluation of heavier precipitation.

\end{document}